\documentclass[twocolumn,secnumarabic,amssymb, nobibnotes, aps, prd]{revtex4}
\usepackage{graphicx}

\begin{document}
\title{Could dissipationless current be observed at non-zero resistance?}
\author{V.L. Gurtovoi, A.I. Ilin, A.V.  Nikulov, and V.A. Tulin}
\affiliation{Institute of Microelectronics Technology and High Purity Materials, Russian Academy of Sciences, 142432 Chernogolovka, Moscow District, RUSSIA.} 
\begin{abstract} The persistent current, i.e. the equilibrium direct electric current circulating in realistic rings, some authors interpreted as dissipationless in spite of non-zero resistance of the rings, whereas the other one suppose that this current can not decay at dissipation. The observation of potential difference connected with the persistent current may give new important information about paradoxical nature of this quantum phenomena observed in nanostructures.
 \end{abstract}

\maketitle

\narrowtext

\section*{Introduction}

The mesoscopic phenomenon predicted as far back as 1970 [1] could be at last observed with confidence [2,3] thanks to nanotechnology progress. The persistent current has been the focus of considerable theoretical and experimental work, but its basic properties remain a topic of controversy. The direct circular current is observed at equilibrium condition in realistic normal metal rings containing atomic defects, grain boundaries, and other kinds of static disorder. The main problem may be put into the question: "How can the persistent current contrive to do not decay in the absence of an applied voltage and a nonzero electrical resistance?" The dictums of different authors concerning a possible answer on this question are contradictory. The authors of [2] and also [4] are sure that a dissipating energy is absent in spite of non-zero resistance and interpret the persistent current as a dissipationless current. Contrary to this confidence I.O. Kulik, who predicted first the persistent current, accentuated in [1] that the taking into account of dissipation should not result in the current disappearance. The experimental results presented in this work testify in favour for the Kulik's opinion. 

 \section {Could familiar analog in atomic physics be complete?}
The author [4] writes that {\it The idea that a normal, nonsuperconducting metal ring can sustain a persistent current - one that flows forever without dissipating energy - seems preposterous}. And it is written in [2] that {\it "A dissipationless equilibrium
current flowing through a resistive circuit is counterintuitive"}. It is correct. The authors [2,4] and all others can not explain how an electric current passing through resistors can be dissipationless. The authors [2] limit oneself to {\it "a familiar analog in atomic physics: Some atomic species' electronic ground states possess nonzero orbital angular momentum, which is equivalent to a current circulating around the atom"}. Indeed, the possibility of the persistent current is described with the Bohr-Sommerfeld quantization 
$$\oint_{l}dl\hbar \nabla \varphi = \oint_{l}dl p = 2\pi \hbar n \eqno{(1)}$$
as well as of the stationary electron orbits of atom. But the wave function $\Psi = |\Psi |\exp{i\varphi }$ describing a electron state in realistic metal ring can not be stationary in time in contrast to the one in atom. Electrons have a finite mean free path in realistic metal ring and the experimental results [2] conform to the theory [5] considering this real case. This experimental result [2] contradict to the statement of their authors that the persistent current, which they observe, is dissipationless.

\section {Persistent current and conventional circular current}
On the other hand I.O. Kulik did not explain how the persistent current can not decay at dissipation without an applied electric field. Conventional circular current $I$ can not decay $RI = -d\Phi /dt$ in a ring with non-zero resistance $R > 0$ only at a non-zero Faraday's electromotive forcee $-d\Phi /dt \neq 0$. The persistent current is observed at a magnetic flux inside the ring $\Phi \neq n\Phi _{0}$, $\Phi \neq (n+0.5)\Phi _{0}$ constant in time $d\Phi /dt = 0$ [2,3]. Its direction and value change periodically in $\Phi $ with period equal the flux quantum $\Phi _{0} = 2\pi \hbar /q$, $I_{p}(\Phi /\Phi _{0})$ because of the Aharonov-Bohm effect, $\hbar \bigtriangledown \varphi = p = mv + qA$. This difference between the persistent current and conventional circular current is obviously the principal motive to interpret the first as the dissipationless current.

\begin{figure}[]
\includegraphics{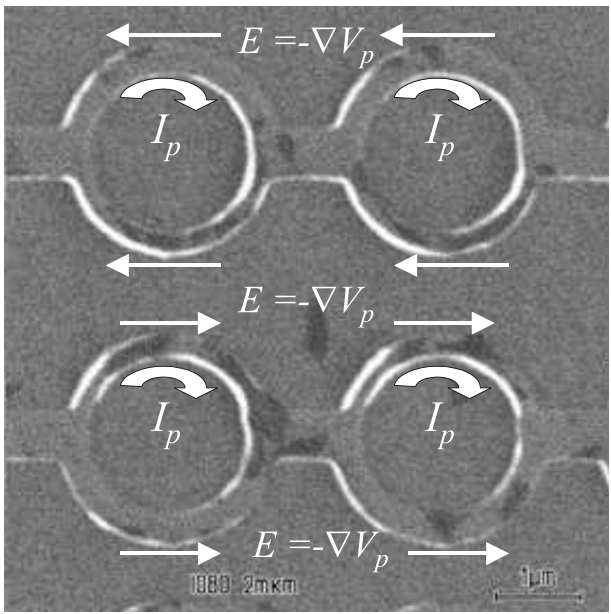}
\caption{\label{fig:epsart} A fragment of the system of 1080 series-connected asymmetric aluminum rings with the same diameter $2r \approx  2 \ \mu m $. The sign of the potential difference $V_{p}$ and the electric field $E = -\nabla V_{p}$ direction depend on the $I_{p}$ direction in the narrow half-ring $w_{n} \approx  0.3 \ \mu m $, having higher resistance $R_{n} > R_{w}$. In the wide half-ring $w_{w} \approx  0.4 \ \mu m $ the persistent current flows against the electric field $E = -\nabla V_{p}$. }
\end{figure}

\section {Persistent current in realistic ring with different resistance of half-rings}
It is well known that the conventional circular current $I$ induced by the Faraday's electromotive force $-d\Phi /dt \neq 0$ should cause the potential difference 
$$V = 0.5(R_{n} - R_{w})I \eqno{(2)}$$ 
on the half-rings with different resistance $R_{n} > R_{w}$. The $I$ direction corresponds to the electric field $E = -\nabla V - dA/dt$ direction in the both half-rings in this case in accordance with the Ohm's law $E = \rho j$. In the case of the persistent current non-potential electric field $-dA/dt$ is absent $-dA/dt = l^{-1} d\Phi/dt = 0$. Experimental investigations of the possibility to observe a potential difference like (2) on the half-rings with different resistance and non-zero persistent current $I_{p} \neq 0$  could have decisive importance for the understanding of the nature of the persistent current.

\begin{figure}
\includegraphics{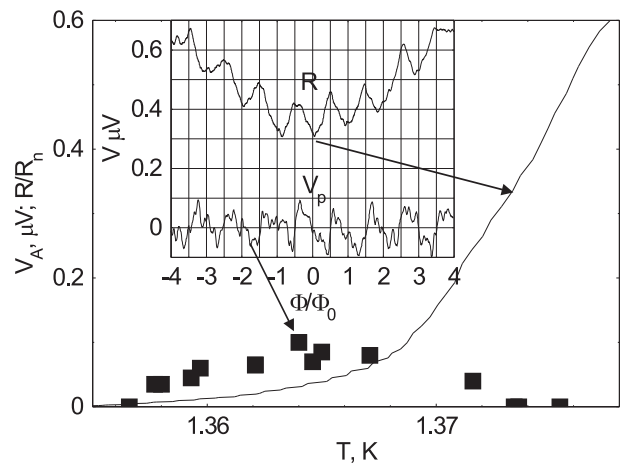}
\caption{\label{fig:epsart} Temperature dependence of the amplitude $V_{A} $ of the potential difference oscillations $V_{p}(\Phi /\Phi _{0}) \propto  I_{p}(\Phi /\Phi _{0})$ (indicated by squares) observed without external electric current on 1080 series-connected asymmetric aluminum rings, shown on the background of the superconducting resistive transition $R/R_{n}$ of this system. The inset shows ($V_{p}$) these oscillations $V_{p}(\Phi /\Phi _{0}) \propto  I_{p}(\Phi /\Phi _{0})$ observed at $T = 1.364 K$ and ($R$) the Little-Parks oscillations of resistance $V = R(\Phi /\Phi _{0})I_{ext} \propto  I_{p}^{2}(\Phi /\Phi _{0})$ observed at $T = 1.374 K$ and very low measuring current $ I_{ext} = 0.1 \ nA$.  }
\end{figure}

Modern nanotechnology allows to make a system of rings, similar the one used in the works [2,3], with different resistance $R_{n} > R_{w}$ of the half-rings. But, as the authors [2,4] note justly, detection of the persistent currents in normal metal ring is extremely difficult, first of all because of its very small magnitude. One may expect that the potential difference (1) should be also very small. The persistent current has much larger magnitude in superconductor ring. Therefore it was detected first at non-zero resistance $R > 0$ as far back as 1962 [6]. The persistent current in normal state of superconductor and non-superconductor (semiconductor and normal metal) has seminar nature and the theorists demonstrate this likeness. I.O. Kulik made the theory of the persistent current in non-superconductor nano-structure [1] just after the work [7] on this phenomenon in normal state of superconductor. In twenty years F. von Oppen and E. K. Riedel have calculated the flux-periodic persistent current in mesoscopic superconducting rings close to $T_{c}$ [8] after the calculation of the disorder-averaged persistent current for a non-superconductor mesoscopic ring [9]. The magnetisation induced by the $ I_{p}(\Phi /\Phi _{0})$ is observed both in normal metal rings [3] and in normal state of superconductor rings [10]. 

In order to verify the possibility of the potential difference $V_{p}(\Phi /\Phi _{0}) \propto  I_{p}(\Phi /\Phi _{0})$ we used a system of 1080 series-connected asymmetric aluminum rings. A fragment of the system is shown on Fig.1. All 1080 rings have the same diameter $2r \approx  2 \ \mu m $ and half-ring widths $w_{n} \approx  0.3 \ \mu m $ and $w_{w} \approx  0.4 \ \mu m $. Without controlled external electric current and depressed uncontrolled electric noise we can observe the voltage oscillations $V_{p}(\Phi /\Phi _{0}) \propto  I_{p}(\Phi /\Phi _{0})$ with amplitude up to $V_{A} \approx  0.1 \ \mu V$ (on one ring $V_{A,1} = V_{A}/1080 \approx  0.1 \ nV$) in the temperature region $T = 1.358 \div 1.372 \ K $ corresponding to the lower part $R(T)/R_{n} = 0.01 \div  0.25$ of superconducting resistive transition, Fig.2, at our possibility to see the  $V_{p}(\Phi /\Phi _{0})$ oscillations with the amplitude down to $V_{A} \approx  0.02 \ \mu V$. The magnetisation measurements [10] and our observation of the Little-Parks [6] oscillations of resistance $R(\Phi /\Phi _{0}) = V(\Phi /\Phi _{0})/I_{ext}$, Fig.2, give evidence of non-zero persistent current at higher temperatures up to $R(T)/R_{n} \approx  1$, in accordance with the fluctuation theory [7,8]. Therefore one may expect that the $V_{p}(\Phi /\Phi _{0})$ oscillations can be also observed at higher temperatures on a system with lager number of the rings.

The author [4] notes that the persistent current can not be detected with help of an ammeter. According to the result of our measurements it can be detected with help of the voltmeter. It is important. The direct voltage can be added in system of series-connected rings. Even if the voltage $V_{p}(\Phi /\Phi _{0}) \propto  I_{p}(\Phi /\Phi _{0})$ is very small in single asymmetric normal metal rings it can be observed with help of a system containing enough big number of rings. The observations $V_{p}(\Phi /\Phi _{0}) \propto  I_{p}(\Phi /\Phi _{0})$ reveal that the persistent current can flow against the force $F_{E} = qE$ of the direct electric field $E = -\nabla V_{p}$, Fig.1. Because of this result the interpretation [2,4] of the persistent current as dissipationless phenomenon can not solve the problem of force balance.   

\section*{Acknowledgement}
This work has been supported by a grant "Possible applications of new mesoscopic quantum effects for making of element basis of quantum computer, nanoelectronics and micro-system technic" of the Fundamental Research Program of ITCS department of RAS and the  Russian Foundation of Basic Research grant 08-02-99042-r-ofi..


\begin{thebibliography}{99}

\bibitem{Tulin1} I.~O.~Kulik, {\em Pisma Zh.Eksp.Teor.Fiz.} {\bf 11}, 407 (1970) ({\em JETP Lett.} {\bf 11}, 275 (1970)).

\bibitem{Tulin2} A. C. Bleszynski-Jayich et al., {\em  Science} {\bf 326} 272 (2009). 

\bibitem{Tulin3} H. Bluhm et al., {\em Phys. Rev. Lett.} {\bf 102}, 136802 (2009)

\bibitem{Tulin4} N. O. Birge, {\em  Science} {\bf 326} 244 (2009).

\bibitem{Tulin5} E.~K.~Riedel and F.~von Oppen, {\em Phys. Rev. B} {\bf 47}, 15449 (1993).

\bibitem{Tulin6} W. A. Little and R. D. Parks, {\em Phys. Rev. Lett.} {\bf 9}, 9 (1962)

\bibitem{Tulin7} I.~O.~Kulik, {\em Zh.Eksp.Teor.Fiz.} {\bf 58}, 2171 (1970)

\bibitem{Tulin8} F.~von Oppen and E.~K.~Riedel, {\em Phys. Rev. B} {\bf 46}, 3203 (1992).

\bibitem{Tulin9} F.~von Oppen and E.~K.~Riedel, {\em Phys. Rev.Lett.} {\bf 66}, 587 (1991).

\bibitem{Tulin10} N. C. Koshnick et al., {\em Science} {\bf 318}, 1440 (2007)

\end{thebibliography}
\end{document}